# Calculation of generalized spin stiffness constant of strongly correlated doped quantum antiferromagnet on two-dimensional lattice and it's application to effective exchange constant for semi-itinerant systems

Suraka Bhattacharjee[*] and Ranjan Chaudhury
Department of Condensed Matter Physics and Material Sciences
S.N. Bose National Centre for Basic Sciences,
Saltlake, Sector-III, Block-JD, Kolkata-700098, India
Email: [*]surakabhatta@bose.res.in
ranjan@bose.res.in

*Abstract*

The generalized spin stiffness constant for a doped quantum antiferromagnet has been investigated both analytically and numerically as a function of doping concentration at zero temperature, based on the strongly correlated t-J model on two-dimensional square lattice. The nature of the theoretical dependence of the stiffness constant on doping shows a striking similarity with that of the effective exchange constant, obtained from the combination of other theoretical and experimental techniques in the low doping region. This correspondence once again establishes that spin stiffness can very well play the role of an effective exchange constant even in the strongly correlated semi-itinerant systems. Our theoretical plot of the stiffness constant against doping concentration in the whole doping region exhibits the various characteristic features like a possible crossover in the higher doping regions and persistence of short range ordering even for very high doping with the complete vanishing of spin stiffness occurring only close to 100% doping. Our results receive very good support from various other theoretical approaches and also brings out a few limitations of some of them. Our detailed analysis highlights the crucial importance of the study of spin stiffness for the proper understanding of magnetic correlations in a semi-itinerant magnetic system described by the strongly correlated t-J model. Moreover, our basic formalism can also be utilized for determination of the effective exchange constant and magnetic correlations for itinerant magnetic systems, in general in a novel way.

## I.   Introduction

The discovery of high temperature superconductivity in the doped layered quantum antiferromagnetic insulators, is a revolutionary phenomenon in the field of both theoretical and experimental condensed matter physics in recent times [1,2]. The relevant antiferromagnetic parental compounds like $La_2CuO_4$ or $YBa_2Cu_3O_6$ can be well understood within the nearest neighbour quantum Heisenberg Hamiltonian, whereas the description of the doped compounds (like $La_{2-x}Sr_xCuO_4$ or $YBa_2Cu_3O_{7+y}$) require quite different type of modeling. Two dimensional t-J model derived from strongly correlated Hubbard model is one such model to explain the magnetic correlations of these doped antiferromagnets. The long range Neel ordering is lost in cuprates at temperature above the corresponding Neel temperatures and these materials can be treated as purely two dimensional systems in this regime. These pure two dimensional systems can be very well treated with t-J model. The inclusion of next-nearest-neighbour interaction has many significances including the pair formation in d-wave superconducting state [3]. Moreover, recently renormalized mean field (RMF) t-J model has been used to study the possibility of d-wave pairing in high temperature superconductors, which is able to produce results comparable to the variational Monte Carlo results [4]. In addition to it, two dimensional t-J model on square lattice based on infinite projected-entangled pair states, has been used to study the occurrences of stripes and the competition of uniform d-wave states versus striped states [5,6].

Another issue that has been much talked about till today is the frustration in the quantum antiferromagnets where the most studied example is the $J_1$-$J_2$ Heisenberg model for spin ½ systems (where $J_1$ and $J_2$ are the nearest and next-nearest-neighbour exchange constants respectively). Sometimes even the third nearest-neighbour exchange interaction $J_3$ is taken into account and phase diagram of $J_1$-$J_2$-$J_3$ Heisenberg model, bearing the signature of possible quantum phase transitions, has been studied [7,8]. The spin fluctuations in doped quantum antiferromagnets is studied based on the extended t-J model taking into account the next-nerarest-neighbour hopping term (t′) [9]. Moreover, the t-$J_1$-$J_2$ model is formulated for investigating the orbital-selective superconducting pairing, gap anisotropy and detailed magnetic behavior of the iron based superconductors like iron pnictides having co-existing localized and itinerant character [10,11].

In this paper, we investigate the magnetic properties of two dimensional doped strongly correlated quantum antiferromagnets using the nearest-neighbour t-J model. This model involves mobile holes, which has a vast applicability in the layered cuprate systems [2,12]. There are also other experiments on the cuprate layers, consistent with the Monte Carlo results, that showed the exponential decay of correlation length and then a power law decay with the increase in doping concentration [13]. Vajk et al. studied the magnetometry of Zn or Mg doped $La_2CuO_4$ and noted the low-temperature tetragonal (LTT) structural phase transition above 10% doping concentration [13]. Above 25% doping the Neel ordering occurs in the LTT phase and persists upto 40.7% site dilution [13]. Previously there were many studies based on the different aspects of the t-J models, some of which are mentioned here. Mori's projection is a powerful technique that establishes a connection between susceptibility and self-energy using the Green's function in relaxation-function theory. The study of two dimensional t-J model using the Mori's projection technique was done in the doping range $0 \leq \delta \leq 0.16$ and the obtained homogeneous solution negates the possibility of stripes formation or phase separation in this doping region [14]. On the contrary our results predict a possible phase separation below $\delta \leq 0.61$ based on the strongly correlated t-J model. Again, the Mori's projection technique is used to investigate the hole and spin excitation spectrum in the background of t-J model [15]. A more complicated scenario is studied applying t-J model in two oppositely doped bi-layers of Mott insulators forming inter-layer excitons and making it difficult to find any single layer analogue [16]. The memory function method has also been used to find the charge dynamics and optical and d.c. conductivity in the t-J model [17,18,19]. Later on, the electron spectrum and superconductive pairing in the t-J model in a paramagnetic state is studied using the projection technique of the two-time Green's function, consistent with the Eliashberg equations [20]. Among the other methods extensively used are the variational Monte Carlo and the Gutzwiller wave function approach minimizing the effective single particle Hamiltonian and structure of gap function [21]. The diagrammatic expansion of Gutzwiller wave function has shown the possibility of superconducting pairing for U/t ≥ 3 and for δ≤ 0.32 [21]. But all the above variational calculations involve only the lowest order contribution in the couplings 't' and 'J'. The calculation of dynamical spin susceptibility for the study of spin dynamics in the t-J model was also done considering only the lowest-order approximation, which has been later proved to be insufficient in correctly determining the dynamical spin susceptibility [18,22]. The self-energy calculation involving the Mori-type technique for two-time thermodynamic Green's function takes into account the spin excitation which is proportional to the square of hopping amplitude ($t^2$) and the imaginary part of self-energy $\Sigma'' \propto t^4$ [23-25]. The consideration of the higher order terms can correctly predict the disappearance of long-range ordering with increase in doping which is in general agreement with our results. In Ref.[24], the critical value of doping concentration ($\delta_c$), above which the long range order vanishes, is also derived for different values of J/t ratio and $\delta_c$ is found to be proportional to J/t. Although these were among the many successful theoretical attempts based on the t-J model with experimentally consistent results, none of them were sufficiently conclusive in predicting the detailed magnetic behaviour of strongly correlated itinerant antiferromagnets, keeping the 'no double occupancy' condition intact. In this context, our present study can elucidate on the complete magnetic behaviour of the doped Mott insulators starting from a first principle calculation based on the t-J model. Regardless of the permanent importance of the t-J model studied so far, it is pertinent to mention here that this model is oversimplified and can be physically applied only to very lightly doped region. The heavily doped region in reality shows substantial weakening of the correlation between the charge carriers and may not be governed by the t-J model in real systems [26].

We calculate here the generalized spin stiffness constant of strongly correlated t-J model based on an idea originally proposed by Kohn and Thouless, both analytically with more rigour and numerically as well and study it in details as a function of doping concentration in the entire doping region [27]. The results are derived using Gutzwiller state and for the strict implementation of the complete 'no double occupancy condition' (NDOC), the variational parameter α is chosen as unity, without carrying out any variational calculation. Our results predict the weakening of spin stiffness constant with increase in doping concentration, which is in general agreement with the results from a previous analytical calculation [28]. This earlier treatment had also predicted the quantum melting of the long range antiferromagnetic ordering at a critical doping concentration, consistent with other theoretical and experimental results from cuprates [28-31].

The temperature dependence of the spin-spin correlation length in two dimensions has been derived analytically by Chakravarty, Halperin and Nelson (CHN) for pure two-dimensional nearest neighbor Quantum Heisenberg Antiferromagnet corresponding to undoped parent compound ($La_2CuO_4$). The calculations have been done using Renormalization Group technique, starting from a field theoretic action [32]. The calculation predicts the decrease of correlation length as temperature increases. This result has been modified later to take into account the large excess of charge carriers in real materials like laboratory-grown $La_2CuO_4$ [33]. Although the formulation of CHN was originally developed for pure (undoped) cuprates, similar results were obtained even in the case of very lightly doped cuprates [34]. This was demonstrated by Manousakis for 'static holes', which is a crude assumption in the sense that completely static holes is impossible in the present scenario. In the very low doping region however, the holes with heavier masses are rather constrained within a small spatial region around the dopant atoms [34,35]. This situation may be regarded as that of nearly static holes. Moreover, in such a low doping region, the hopping amplitude is negligibly small.

The neutron scattering experimental results of Thurston et al. show the fall of 2-dimensional correlation length ($\varepsilon_{2D}$) with increase in doping concentration (x) of $La_{2-x}Sr_xCuO_4$ [2]. On the other hand, our numerical results for spin stiffness constant qualitatively and even quantitatively describe the nature of its decrease with increase in doping starting from the half-filled band limit and interestingly is quite similar to the behaviour of the result obtained from the above experimental work. This carries a strong hint that spin stiffness constant may very well play the role of an effective exchange constant in doped antiferromagnets. In addition, our analytical calculations and nature of our calculated spin stiffness constant are in qualitative agreement with the Quantum Monte Carlo results, which also predicts the reduction of antiferromagnetic ordering with the increase in doping [2,34].

A slightly different type of concept and estimate for effective exchange constant in an effective ideal Fermi-sea like background was introduced by Himeda and Ogata and we have presented a detailed comparison between their results and ours as well [29,30]. The vanishing of our spin stiffness constant as $\delta \rightarrow 1$ agrees with the results of Himeda and Ogata; however their calculations did not produce any plausible signature of phase transition which comes as a possible outcome of our analytical and numerical calculations.

It is also quite important to keep in mind that the doped phase of quantum Heisenberg antiferromagnet described by the t-J model, behaves essentially as an itinerant magnetic system with strong on-site correlations because of the presence of mobile holes with spin. Therefore our effective exchange constant is to be looked upon as that appropriate to an itinerant magnet. In this context it is worthwhile mentioning that, the previous researches for determining the magnetic interactions done using the density-functional theory, were for completely localized spins [36-38]. There were a few attempts to study the properties of conventional itinerant magnetic systems using coherent-potential, where local exchange-correlation approximation was used in the band calculations [39]. Later Antropov calculated the effective exchange coupling of the itinerant systems using a combination of 'inverse susceptibility' approach and multiple-scattering theory [40,41]. However, the above techniques and calculations were rather insufficient to determine the effective exchange coupling and magnetic interactions in the itinerant systems in the presence of strong Coulomb correlations. Henceforth, our rigorous analytical and numerical calculations presented here provide a simple and straight forward way for determination of effective exchange coupling and description of the magnetic correlations of strongly correlated semi-itinerant systems both qualitatively and quantitatively.

Thus, one of the major aims of this paper is to theoretically determine and establish a more quantitative relation between generalized spin stiffness constant and effective exchange constant corresponding to a doped strongly correlated quantum antiferromagnet in 2-dimension, described by the nearest-neighbour t-J model, and then use it to study effective exchange constant as a function of doping concentration from first principle.

It may be recalled that the equivalence of stiffness constant with effective coupling is well known in many other problems such as Berezinsky-Kosterlitz-Thouless (BKT) transition [42, 43].

## II. Mathematical Formulation and Calculation

The Hamiltonian of the strongly correlated nearest neighbour t-J model is given by [44]

$$H_{t-J} = H_t + H_J \tag{1}$$

where

$$H_t = \sum_{<i,j>,\sigma} t_{ij} X_i^{\sigma 0} X_j^{0\sigma} \tag{2}$$

and

$$H_J = \sum_{<i,j>} \{J_{ij} (S_i \cdot S_j - (\tfrac{1}{4}) n_i n_j)\} \quad \text{(with } J_{ij} > 0\text{)} \tag{3}$$

where $t_{ij}$ is the nearest-neighbour hopping amplitude connecting $j^{th}$ and $i^{th}$ site and $J_{ij}$ is the exchange constant between the carriers on nearest neighbours; X's are the Hubbard operators, satisfying the appropriate commutation relations and the usual Hubbard algebra [28]. For nearest neighbour hopping and exchange interaction we take $t_{ij} = t$ and $J_{ij} = J$. It may be recalled that the quantities 't' and 'J' are considered independent.

The generalized spin stiffness constant $\vec{D}_{spin}$ is defined as [28],

$$\vec{D}_{spin} = \lim_{\phi \to 0} (\tfrac{1}{2}) \frac{\delta^2 E}{\delta \phi^2} \tag{4}$$

where $E(\phi)$ is the total ground state energy in the presence of staggered Peierl's phase (resembling a magnetic flux) $\phi_\sigma$ with $\sigma = \uparrow$ or $\downarrow$, arising from an applied vector potential $A(r)$, such that

$$\phi_\downarrow = -\phi_\uparrow = \phi \tag{5}$$

The hopping amplitude $t_{ij}$ for a fermion with spin $\sigma$ is modified to $t_{ij} e^{i\phi_\sigma}$, only if the vector potential has a component along the direction of hopping. We have included the factor of ($\frac{e}{\hbar c}$) in the phase $\phi$ in the final expression of $\vec{D}_{spin}$ with proper scaling.

The total spin stiffness constant $\vec{D}_{spin}$, abbreviated as '$\vec{D}_s$', may be written as,

$$\vec{D}_s = \vec{D}_s^t + \vec{D}_s^J \tag{6}$$

where $\vec{D}_s^t$ and $\vec{D}_s^J$ are the contributions from the 't' term and the 'J' term respectively. They are defined as

$$\vec{D}_s^t = \lim_{\phi \to 0} (\tfrac{1}{2}) \frac{\delta^2 T}{\delta \phi^2} \tag{7}$$

and

$$\vec{D}_s^J = \lim_{\phi \to 0} (\tfrac{1}{2}) \frac{\delta^2 E_J}{\delta \phi^2} = \lim_{\phi \to 0} (\tfrac{1}{2}) \frac{\delta^2 E_J^{sf}}{\delta \phi^2} \tag{8}$$

where T is the kinetic energy contribution, $E_J$ is the total exchange energy and $E_J^{sf}$ is the spin flip part of the exchange energy, which again are the ground state expectation values of the corresponding parts of the Hamiltonian. It may be pointed out that the direct part of the exchange energy term does not contribute to $\vec{D}_s^J$ [28]. Furthermore, it may be noted that $\vec{D}_s^t$ and $\vec{D}_s^J$ both have the dimension of energy since ø is a dimensionless quantity.

In calculating E, avoiding the rather complicated Hubbard algebra, we make use of the Gutzwiller state with strictly NDOC imposed upon it [45].

The very general form of the Gutzwiller state is given by [45]:

$$|\Psi_G\rangle = \prod_l (1 - \alpha \hat{n}_{l\uparrow} \hat{n}_{l\downarrow}) |FS\rangle \qquad (9)$$

where |FS> is the Fermi sea ground state and α is the variational parameter determined by minimizing the expression for E in the general case. In the case of NDOC however, we have straight away taken α=1, without going into any variational scheme to determine 'α'. As stated earlier, this is in the spirit of the very strong correlation situation assumed to persist even in the doped phase, with effective on-site Coulomb repulsion much larger than the band-width, leading to strict avoidance of double occupancy on each site.

Now expressing the Fermi sea ground state $|FS\rangle$ in terms of fermion creation operators, equation (9) takes the following form :-

$$|\Psi_G\rangle_{\text{NDOC}} = \prod_l (1 - \hat{n}_{l\uparrow} \hat{n}_{l\downarrow}) \prod_{k,\sigma} \sum_{i,j} C^+_{i\sigma} C^+_{j-\sigma}\, e^{i(r_i - r_j)\cdot k} |vac\rangle \qquad (10)$$

where |vac> is the vacuum state (having fermionic occupation number equal to zero at all sites) and we have omitted the normalization constant for the time being which will be included later in the calculation for the energy eigen values. The symbols i, j and l all denote the lattice sites and **k** represents the wave vector for the fermion, bounded by $k_F$ (Fermi wave vector) from above [28]. Here $k_F$ is defined corresponding to the sea of holes which we have considered as the existing carriers in the presence of doping (vacancies). It may be recalled that the insulating phase corresponds to one hole per site with the holes being immobile.

The Fermi wave vector for the two dimensional systems,
$$k_F = \sqrt{2\pi n}/a \qquad (11)$$

where 'n' is the concentration of existing fermionic carriers (holes) present in the doped system and 'a' is the lattice spacing. 'n' is related to the doping (vacancy) concentration 'δ'(doping introduces vacancies in the system by removing carriers) as,
$$n = 1 - \delta \qquad (12)$$

Combining equations (11) and (12), $k_F$ gets related to δ as,

$$k_F = \sqrt{2\pi(1-\delta)}/a \qquad (13)$$

The equation (3) can now be rewritten as:

$$H_J = \sum_{i,j} J_{ij}\, H'_J \qquad (14)$$

where
$$H'_J = S_i \cdot S_j - \left(\frac{1}{4}\right) n_i n_j \qquad (15)$$

Again,
$$E_J = \left(\frac{4 t_{\text{eff}}^2 \cos(2\emptyset)}{V_{\text{eff}}}\right) \frac{{}_{\text{NDOC}}\langle \Psi_G | H'_J | \Psi_G \rangle_{\text{NDOC}}}{{}_{\text{NDOC}}\langle \Psi_G | \Psi_G \rangle_{\text{NDOC}}} \qquad (16)$$

where $_{NDOC}\langle\Psi_G|\Psi_G\rangle_{NDOC}$ is the normalization for the Gutzwiller state; $t_{eff}$ is the effective nearest neighbour hopping amplitude and $V_{eff}$ is the effective on-site Coulomb barrier potential in the doped phase for infinitesimal doping ie., $\delta \to 0$ within the effective one band scenario [28]. Thus we can model the initial J ($J_{bare}$ or $J(\delta)$ with $\delta \to 0$) as $4t_{eff}^2/V_{eff}$ as considered in the t-J model and estimated the initial t/J ratio (in the limit $\delta \to 0$) as $V_{eff}/4t_{eff}$ [28].

Carrying out detailed and more rigorous calculation after normalizing the Gutzwiller state and considering only the contribution from the nearest neighbour interaction we get the following result for the exchange energy contribution (the major steps of the calculation are shown in Appendix(A)) :

$$\frac{_{NDOC}\langle\Psi_G|H'_J|\Psi_G\rangle_{NDOC}}{_{NDOC}\langle\Psi_G|\Psi_G\rangle_{NDOC}} = \prod_k^{k_F} 2(1-\delta)^2 \qquad (17)$$

The square on $(1-\delta)$ is the consequence of the exchange interaction operating only between the occupied sites. Making use of equation (8) and by taking derivative of equation (16) twice we get the exchange part of the spin stiffness constant as,

$$\tilde{D}_s^J = -4J \prod_k^{k_F} 2(1-\delta)^2 \qquad (18)$$

From the earlier calculation described in [28], the kinetic energy contribution of the Fermionic system at zero temperature can be found from the following equation,

$$T = \frac{_{NDOC}\langle\Psi_G|H_t|\Psi_G\rangle_{NDOC}}{_{NDOC}\langle\Psi_G|\Psi_G\rangle_{NDOC}} \qquad (19)$$

The above quantity is evaluated in the presence of staggered phase $\phi_\sigma$ (staggered phase corresponding to up or down spins) making use of the orthogonality of the independent states and the result comes out to be (the major steps of the calculation are shown in Appendix(B)):

$$T(\phi \neq 0) = (t)[\prod_{k,\sigma}^{k_F} \sum_\sigma 4\cos(ka)(1-\delta)^2 \cos(\phi_\sigma) - N_l \prod_{k,\sigma}^{k_F} \sum_\sigma 4\cos(ka)\cos(\phi_\sigma)/N^2] \qquad (20)$$

Here we use equation (5) for $\phi_\sigma$ corresponding to up and down spin respectively; '$N_l$' is the expectation value of the number operator corresponding to the total number of lattice sites singly occupied by spins corresponding to the holes and 'N' is the total number of sites.

Thus, $\qquad N_l = N(1-\delta) \qquad (21)$

For 2D lattice, the vector potential A(**r**) is assumed to be applied along the x direction and making use of equations (7) and (20), we get the expression for the kinetic part of the spin stiffness constant as:

$$\tilde{D}_s^t = (-t)[\prod_{k,\sigma}^{k_F} 4\cos(k_x a)(1-\delta)^2 - N_l \prod_{k,\sigma}^{k_F} 4\cos(k_x a)/N^2] \qquad (22)$$

The second term in equation (22) is physically important since it signifies the complete projecting out of double occupancy in the occupied sites $N_l$. The increase in doping concentration '$\delta$' decreases the number of occupied sites, thus decreasing the probability of double occupancy.

The vanishing of the total spin stiffness constant $\vec{D}_s$ implies the loss of rigidity (rigidity arising from the antiferromagnetic coupling) of the spins of the carriers (holes) in the doped phase.

Again the vanishing of $\vec{D}_s^t$ can arise from the vanishing of $\cos(k_x a)$ at $k_x = \pi/2$ and for the whole set of values of $k_x$ ($0 \leq |k_x| \leq k_F$), at least one value should satisfy the above relation. Hence the boundary condition for the vanishing of $\vec{D}_s^t$ should be determined by $k_F = \pi/2$.

Thus from equation (11) and (13),

$$k_F = \sqrt{2\pi(1-\delta)}/a = \sqrt{2\pi n}/a = \pi/2 \qquad (23)$$

This condition leads to 
$$n \geq 0.39 \qquad (24)$$
$$\text{ie., } \delta = (1-n) \leq 0.61 \qquad (25)$$

The above inequality is the same as was obtained earlier [28]. So for doping concentration less than 0.61, $\vec{D}_s^t$ goes to zero. Therefore the region below 61% doping is entirely governed by spin stiffness from the exchange part ($\vec{D}_s^J$). Again $\vec{D}_s^J$ vanishes only when $\delta \to 1$ ie, for 100% doping and for $\delta \to 1$, the concentration of hole carriers (n) vanishes resulting in the vanishing of $k_F$ and $\vec{D}_s^t$ as well. Hence the total spin stiffness constant falls with increasing doping concentration and exactly goes to zero for $\delta = 1$. Our detailed numerical results elaborated later show that the stiffness constant practically vanishes at a much lower value of doping concentration, but a negligibly small value prevails and it theoretically tends to zero as $\delta \to 1$. This result is in quantitative agreement with that of Himeda and Ogata that antiferromagnetic correlation prevails upto 100% doping [29,30].

Let us now come back to the conjecture involving the relation between the spin stiffness constant and the effective antiferromagnetic exchange coupling between mobile holes in the doped phase, as stated earlier. In order to test this conjecture, total spin stiffness constant is first of all scaled down by the effective number of pair of holes, $^{N_l}C_2$, where $N_l$ has been defined earlier. This makes the comparison between the spin stiffness constant and the antiferromagnetic exchange constant more meaningful and transparent in the background of a semi itinerant magnetic system produced by doping. Moreover, $\vec{D}_s$ shows a very drastic fall with very small increase of $\delta$ and in contrast to it, the scaled stiffness constant shows a comparatively moderate fall with the increase of $\delta$, which is much more alike to the plot obtained from the combined results of experiments and Monte Carlo calculations. We have verified this result for all the lattice sizes including the 200x200 lattice, the largest lattice size we could handle here.

Thus the total spin stiffness constant corresponding to a single pair of mobile holes to be denoted as '$D_s$' is given as:

$$D_s = (\vec{D}_s^J + \vec{D}_s^t)/{^{N_l}C_2} \qquad (26)$$

This new quantity $D_s$ is then calculated from our earlier obtained results for $\vec{D}_s$ with parameters appropriate to $La_{2-x}Sr_xCuO_4$ for different lattice sizes by making use of equations (18) and (22) for enumeration and is plotted against doping concentration ($\delta$). Our theoretical graph is then compared with the experimental results in combination with those from other theoretical and computational techniques, as will be discussed in the next section.

## III. Calculational Results and Comparison with Phenomenology and other Theoretical Approaches

Let us first of all review the relevant experimental and other theoretical and computational results for this problem. Neutron scattering studies have been carried out on $La_{2-x}Sr_xCuO_4$ samples at different doping concentrations. The results reveal the presence of finite intraplane magnetic correlation in 2-dimension above it's Neel temperature ($T_N$=190K) [2]. Above $T_N$, the long range interplane correlation is lost and 2-dimensional correlation length in pure $La_2CuO_4$ is ~ 200 A° at 300 K. But in this temperature range the planes are still at low temperature since T << intraplanar J.

The 2D antiferromagnetic correlation length, $\varepsilon_{2D}$, has been measured in double-axis (energy integrating) experiments on a number of doped samples. It was found that $\varepsilon_{2D}$ is approximately independent of temperature, but it strongly depends on doping concentration [2] (see Fig.(1)).

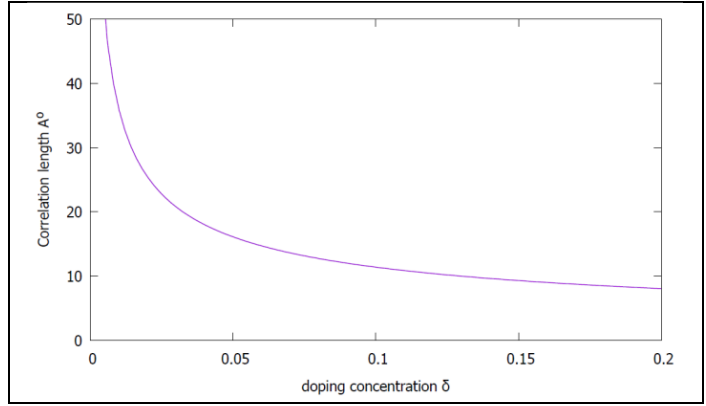

Fig.(1). Magnetic correlation length vs. doping concentration 'δ' of $La_{2-x}Sr_xCuO_4$

This purely experimental plot gives the relation between 2-dimensional correlation length and doping concentration. However to extract the dependence of effective exchange constant on doping concentration from this, the Monte Carlo results for 2-dimensional Quantum Heisenberg Antiferromagnetic Model (2D QHAFM) as will be discussed below. It must however be stressed that this combined semi-phenomenological scheme is only to provide a link between the results from our rigorous theoretical approach and the experimental situation.

CHN calculated the temperature dependence of the magnetic correlation length of the spin ½ Heisenberg antiferromagnet corresponding to pure $La_2CuO_4$ using renormalization group analysis of quantum non-linear σ model (QNLσM). They obtained for T→0 [32]:

$$\varepsilon_{2D} = C_\varepsilon \exp\left[\frac{2\pi\rho_s}{K_B T}\right] \qquad (27)$$

where $\varepsilon_{2D}$ is the 2-dimensional antiferromagnetic correlation length and $2\pi\rho_s$ (=1.25J for undoped state corresponding to half-filled band) is the well known spin wave stiffness constant which is proportional to the bare nearest neighbour antiferromagnetic exchange constant 'J'. However, the significance of our derived generalized spin stiffness constant for the doped Heisenberg antiferromagnet is quite different from the spin wave stiffness constant defined above. Hence it is apprehended that the spin wave stiffness constant, which is also proportional to $J_{bare}$, should be different in magnitude from our derived doping dependent generalized spin stiffness constant. Moreover, the applicability of the t-J model is restricted to slightly less than half-filled band limit in the low doping side, which constraints the carrying out of the calculations at exactly $\delta=0$. Hence in our formalism all the calculations are restricted to $\delta\rightarrow 0$ limit.

Again Quantum Monte Carlo (QMC) studies have been performed on 2-dimensional Heisenberg antiferromagnets by Manousakis for very low doping at $T\rightarrow 0$. The holes in such a low doping limit are considered to be almost localized and the increase in doping enhances the itinerancy in the system. This character of doping is also experimentally observed in the layers of Sr doped $La_2CuO_4$ [46]. The numerical results of Manousakis in this 'nearly static hole regime' are fitted quite well with a function of exponential form, as in equation (27). The best fit is given by [34]:

$$\xi_{2D} = \left(\frac{0.276a}{\sqrt{(1-\delta)}}\right) \exp[1.25\, J/T] \qquad (28)$$

where $a=3.77 A^o$ is the lattice constant for $La_2CuO_4$ and $0.276a$ is the prefactor for pure 2-dimensional Heisenberg antiferromagnet with 'J' appearing in the above equation to be regarded as '$J_{eff}(\delta)$' [47].

On the other hand the best fit of the experimental result from neutron scattering, neglecting the weak temperature dependence, is found to be [2],

$$\xi_{2D} \approx \frac{3.8}{\sqrt{\delta}} \qquad (29)$$

Combining equations (28) and (29), one arrives at the following semi-phenomenological relation between the effective antiferromagnetic exchange constant $J_{eff}$ (>0) and $\delta$ at in the very low doping regime (neglecting the temperature dependent prefactor),

$$J_{eff} \sim -\ln\left[0.075\left(\frac{\delta}{1-\delta}\right)\right] \qquad (30)$$

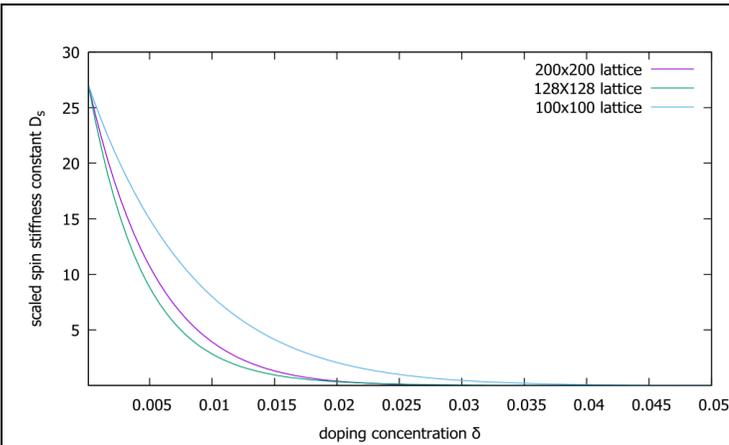

Fig.(2a). Scaled spin stiffness constant '$D_S$' vs. doping concentration '$\delta$' plot obtained from our analytical calculation using t-J model with $t \sim 8J$ for three different lattice sizes (100x100,128x128,200x200)

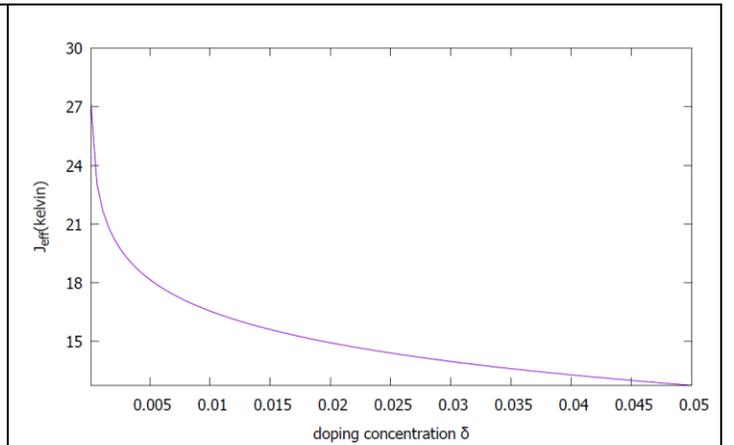

Fig.(2b). $J_{eff}$ vs. doping concentration '$\delta$' plot using equation (30)

It is very clear from the strong similarities between the nature of the graphs seen in the two plots in Fig.(2a) and Fig.(2b), that the pair spin stiffness constant ($D_s$), calculated by us from the t-J model with strict NDOC at zero temperature and multiplied with proper proportionality constant, can very well represent the real physical effective antiferromagnetic exchange coupling constant '$J_{eff}$' in the presence of doping at least qualitatively. Hence, the pair spin stiffness constant can truly be considered as equivalent to the effective exchange coupling at least in the very low doping region. Here it may be noted that the plots show the effective exchange constant starting from δ→0 limit (slightly less than half-filling) as the t-J model is not valid at exactly δ=0. The similarity is also prevalent for other samples with different band widths (2t) and Coulomb repulsion barrier $V_{eff}$ ie., corresponding to different initial t/J ratios.

Having established the equivalence of our calculated our calculated '$D_s$' and the physical '$J_{eff}$', we now study the variation of our calculated '$D_s$' with doping concentration δ in the entire doping regime (see Figs.3(a)-3(c)). However, it must again be emphasized that the t-J model provides a genuine description of the real doped quantum antiferromagnet only in the low doping regime.

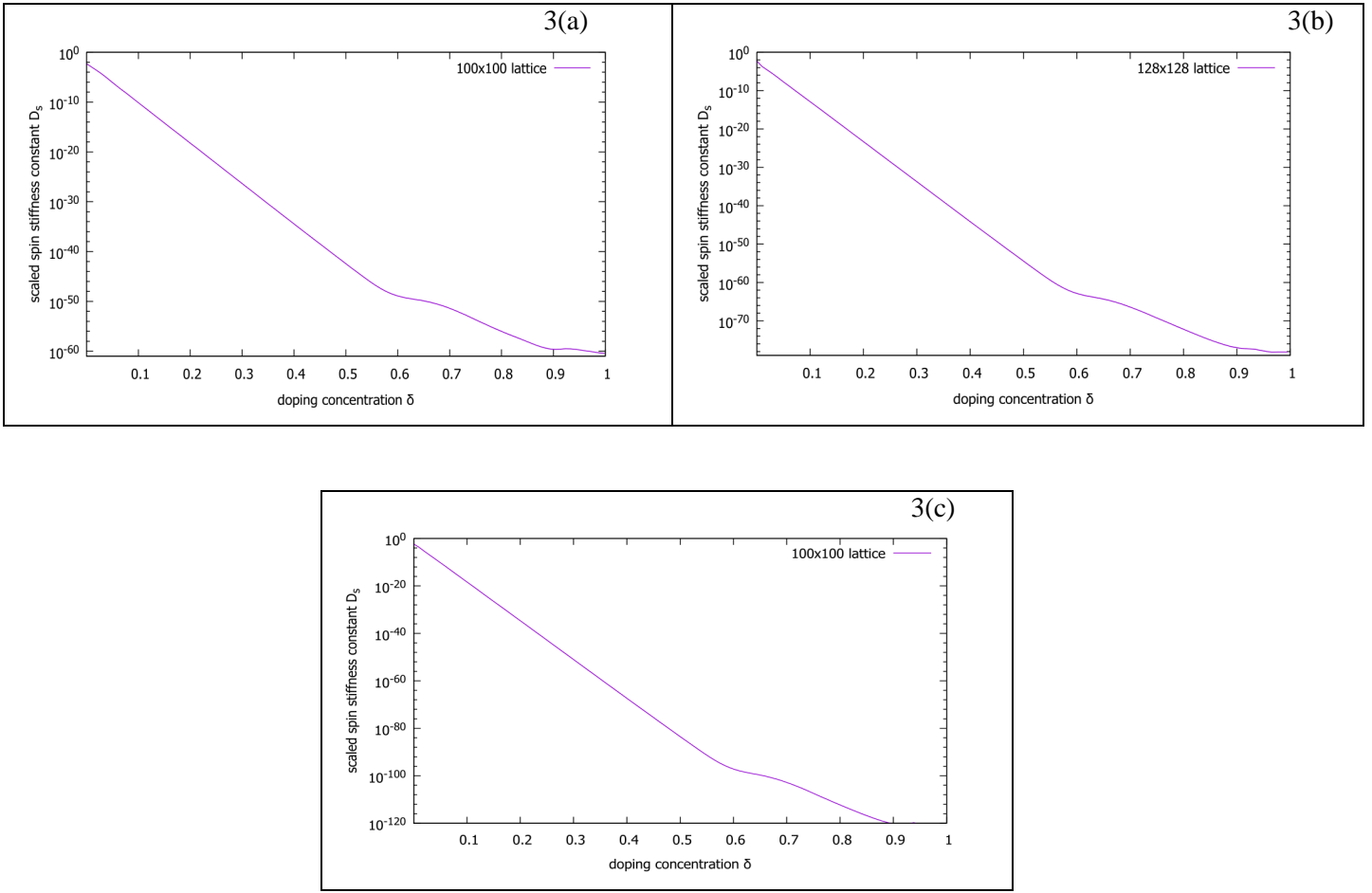

Fig.(3). Scaled spin stiffness constant '$D_s$' vs. doping concentration 'δ' upto 100% doping in logscale;(a)for 100x100 lattice,(b) for 128x128 lattice,(c) for 200x200 lattice

From these plots of ours it can be noticed that there is a huge decrease in the magnitude of spin stiffness constant 'D$_s$' with increase in doping concentration and it practically becomes vanishingly small above 20% doping, which is again supported by the experimental results of Thurston et al. [2]. It can also be noticed that all the curves show the same trend.

It must be highlighted that, some shoulder-like structures or a point of inflexion are seen in all the above plots (Fig. 3(a)-3(c)) near δ=0.61. This may very well indicate a point of cross-over or phase transition at zero temperature. Emery et al. have shown the existence of phase separation in t-J model by both analytical calculation and exact numerical diagonalization in small in finite lattice [48]. The antiferromagnetic phase gets separated into a hole-rich phase and a hole-deficient phase for all J/t ratios below a minimum value of vacancy concentration given by δ < δ$_m$ [48]. For strong correlation (small J/t), phase separation occurs due to kinetic energy frustration in two dimension and the holes are put in a separate region to reduce kinetic energy [48]. The existence of the point of inflexion near δ=0.61 in our analytical and numerical results, which is the artefact of the vanishing of $\widetilde{D}_s^t$ below δ ≤ 0.61, can quite logically represent the phenomenon of phase separation below δ=0.61 as proposed in [48] and absence of it above this doping concentration. The vanishing of the contribution from the kinetic energy part is a plausible signature for the phase separation in this regime where the contribution to the spin stiffness constant arises solely from the exchange interaction term.

Furthermore, the exact vanishing of D$_s$ as δ→1 in our calculational results, is the signature of the persistence of exchange coupling in the form of short range antiferromagnetic ordering almost upto 100% doping. This is quantitatively in agreement with results of Himeda and Ogata obtained from simplified variational calculations as discussed previously [29,30].

Himeda and Ogata attempted to implement the effect of NDOC through the projection operator of Gutzwiller state by renormalizing the magnitudes of hopping amplitude and exchange constant with doping dependent multiplicative factors g$_t$ and g$_J$ respectively in the background of an ideal Fermi sea. These renormalized parameters corresponding to hopping and exchange are related to the un-renormalized ones by the equations,

$$\widetilde{t}_{\text{eff}} = g_t t , \qquad \widetilde{J}_{\text{eff}} = g_J J \qquad (31)$$

where g$_t$ and g$_J$ are the Gutzwiller factors calculated from variational energy calculation by Ogawa et al. [49]. The Gutzwiller factor acts as the enhancement factor on J that introduces the effects of the projection operator in the Gutzwiller state and the z-component of g$_J$ stabilizes the antiferromagnetic order, as shown using the variational Monte Carlo results [29]. The Gutzwiller factor g$_J$ has been derived as [30]:

$$g_J = \frac{4(1-\delta)^2}{(1-\delta^2+4m^2)^2} \qquad (32)$$

where m is the expectation value of antiferromagnetic order parameter denoting the staggered magnetization in the long range antiferromagnetically ordered state. The vanishing of g$_J$ and hence that of $\widetilde{J}_{\text{eff}}$ at δ=1 is clear from the above expression for g$_J$. The quantity m is non-zero at δ→0 and is doping dependent ie., (m = m(δ)). Further m decreases with increasing doping δ and takes a very small value beyond δ≈0.1 [29].

Thus from equation (32), it follows that $g_J$ decreases with increasing δ and approaches zero value as δ→1. This is qualitatively very similar to the fall observed in our $D_s$ vs. δ plot and to the aquiring of vanishingly small values of $D_s$ as δ approaches 1. Very importantly however, Himeda and Ogata could not identify any point of possible phase separation which is present in our plot.

## IV.    Discussion

Our analytical calculations described above show that our calculated $D_s$ theoretically goes to zero at 100% doping concentration. The part of $D_s$ due to kinetic energy part remains zero upto δ=0.61 for 2-dimensional lattices, then increases and again goes to zero at δ=1; however contribution to $D_s$ from the exchange part monotonically decreases from very low δ and vanishes at δ=1. Total $D_s$ is plotted against δ and it bears a striking similarity with the $J_{eff}$ versus δ plot obtained by a combination of QMC results and experimental data in the low doping regime viz. δ≤0.05 (see Figs.2(a) 2(b)) and [34]. We have used the results from Monte Carlo calculation with it's error limitations and experimental results to extract the dependence of physical $J_{eff}$ on δ. Hence these errors and limitations are embedded in the $J_{eff}$ vs δ plot in Fig.(2b). It should be emphasized here however that even in this region of very small doping, the holes are not absolutely static as discussed above, but the kinetic energy contribution itself is very small which allows one to consider the holes as almost static in this region for the QMC based treatment [34,35]. Regarding our results displayed in Fig.(2a), we could perform calculations on maximum lattice size only upto 200x200 for our calculations, which is much below the thermodynamic limit. Moreover, the validity of the t-J model is restricted to δ→0 limit, which restrained us from calculating the spin stiffness constant at δ=0. Despite all these limitations and crudeness, the similarity of the two plots under Fig.(2) is highly significant.

Our calculational results for $D_s$ agree qualitatively with those Himeda and Ogata at very high δ, as discussed previously. The notable absence of any point of inflexion in the mid-high δ regime in the work of latter, in disagreement with our results (see Figs. 3(a)-3(c)), is probably due to inadequate handling of correlation [29,30,49]. On the other hand, in the very high δ regime the system effectively goes over to a weakly correlated phase even with a given repulsive potential [26]. Therefore, it is not surprising that our results agree with those of Himeda and Ogata in this regime.

The detailed band structure effects and interlayer processes are completely neglected in our calculation, although it can help determining the magnetic phase boundaries in some of the real cuprate systems. The transition from long range to short range ordered phase has been studied previously using spin diffusion coefficient calculations [28]. This calculation based on time-dependent Hartree-Fock treatment for dynamical spin susceptibility, showed the survival of the long range ordered phase upto a critical doping concentration of 14% [28]. This implied the existence of a finite value of $T_N$ in the regime 0 ≤δ≤ 0.14, if the system is made quasi-two dimensional [28]. Later more accurate calculations were performed with the higher orders terms in the hopping amplitude 't' viz. ($t^2$ and $t^4$) taken into account in the self energy calculation [24,25]. The fall of correlation length and decay of long range ordered phase with increasing δ for different J/t ratios were studied [24]. According to the above results, the disappearance of long range order at T=0 is predicted from the vanishing of staggered magnetization 'm' at a critical doping concentration $δ_c$ ≈ 0.025 for very small J/t ratio (J/t = 0.2) [24]. This is in agreement with our present numerical results too (Fig.1(a)) [25]. In the present paper based on non-perturbative method, the calculationally obtained sharp decay of $D_s$ for $\delta \leq 0.03$ corresponding to $J/t = 0.125$ does also represent the rapid fall of both $T_N$ and long range order with doping in this regime. This is very well supported by the above result [24]. Besides, one also confirms the existence of "novel paramagnetic phase" from our present calculations, if bare 'J' is taken to be vanishingly small in the t-J model [28].

The highly doped regions in the plots (Figs.3(a)-3(c)) represent in a way the weakly correlated regimes for the system [26,29,20]. The system in this regime appears reasonable to be described by the FL Theory. However, the stringent NDOC at each site ensures the manifestation of the non-Fermi Liquid character of this phase. Our detailed calculations here have been done taking into account only the nearest-neighbour exchange constant 'J' and hopping parameter 't' which automatically leads to the renormalized effective exchange constant '$J_{eff}$' (equivalent to $D_s$). This is in contrast to the various heuristic phenomenological models like, t-t'-t"-J or t-$J_1$-$J_2$ model which also try to understand the doped phase [9,10]. Nevertheless, our first principle approach can very well capture the physics of doped quantum antiferromagnets.

In conclusion, the generalized stiffness constant calculation of ours for the strongly correlated t-J model is quite powerful and does bring out the concept of effective antiferromagnetic exchange constant appropriate to a semi-itinerant system, quite neatly. Furthermore, our theoretical results are in excellent agreement with those from various other theoretical approaches and also brings out limitations of some of them. As stated earlier, the effective exchange constants of some itinerant magnets like Fe, Ni, Gd have been determined using the techniques based on 'inverse susceptibility'. Moreover, the exchange correlation in itinerant magnets can be expressed in terms of the elements of scattering path matrix in the framework of density functional approach [37,40,41]. Band structure calculation based on multiple-scattering theory and spin-spiral techniques has also been done for estimating the exchange interaction in these itinerant magnets [41]. The effective exchange constant involving the nearest neighbour spins is related to the second order derivative of the magnetic energy with respect to the spin fields. Making use of this formalism the effective exchange constant turns out to be the inverse dynamic magnetic susceptibility (DMS), with some assumptions for the weakly interacting systems [40]. All these approaches described above, show that there has been ongoing theoretical research to determine the exchange constant and study the short range correlations even in weakly correlated itinerant magnetic systems, which is still a challenging problem in the field of condensed matter physics. In this context, our scheme based on generalized spin stiffness calculation provides a novel formalism for calculating the effective exchange constant of itinerant magnets, both weakly and strongly correlated.

## V.    *Future plan*

Our future plan includes:
(1) Extension of our present formalism to investigate charge stiffness for t-J model on 2-dimensional lattice.
In combination with our results obtained for spin stiffness, this would help in characterizing the microscopic state of the doped phase of 2-dimensional strongly correlated quantum Heisenberg antiferromagnets more clearly.

(2) Calculation of both spin and charge stiffness for t-J model on 1D lattice.

## Appendix A:

$$H'_J = S_i \cdot S_j - \left(\frac{1}{4}\right) n_i n_j \tag{A1}$$

In terms of fermion operators,

$$H'_J = \sum_\sigma C^+_{i\sigma} C_{i-\sigma} C^+_{j-\sigma} C_{j\sigma} \tag{A2}$$

Thus making use of equation (10),

$$H'_J \, |\Psi_G\rangle_{\text{NDOC}} = \sum_\sigma C^+_{i-\sigma} C_{i\sigma} C^+_{j\sigma} C_{j-\sigma} \prod_l (1 - \hat{n}_{l\uparrow}\hat{n}_{l\downarrow}) \prod_k^{k_F} \sum_{i',j'} e^{i(r_{i'}-r_{j'})\cdot k} |i'\sigma, j'-\sigma\rangle \tag{A3}$$

where '$i'\sigma$' and '$j'-\sigma$' denotes fermions at sites '$i'$' and '$j'$' with spins '$\sigma$' and '$-\sigma$' respectively.

$$H'_J \, |\Psi_G\rangle_{\text{NDOC}} = \prod_l (1 - \hat{n}_{l\uparrow}\hat{n}_{l\downarrow}) \prod_k^{k_F} \sum_{i',j'} \delta_{ii'}\delta_{jj'} e^{i(r_{i'}-r_{j'})\cdot k} |i'-\sigma, j'\sigma\rangle \tag{A4}$$

$$_{\text{NDOC}}\langle\Psi_G|H'_J|\Psi_G\rangle_{\text{NDOC}} = {}_{\text{NDOC}}\langle\Psi_G| \prod_k^{k_F} e^{i(r_i-r_j)\cdot k} |i-\sigma, j\sigma\rangle \tag{A5}$$

Here '$\hat{n}_{l\uparrow}\hat{n}_{l\downarrow}$' does not contribute since exchange is not possible between up and down spins on the same site and i and j are the nearest neighbour occupied sites.

Further,

$$_{\text{NDOC}}\langle\Psi_G|\Psi_G\rangle_{\text{NDOC}} = \langle vac| \prod_{k'} \sum_{i',j',\sigma} e^{-i(r_{i'}-r_{j'})\cdot k'} C_{j',-\sigma} C_{i',\sigma} (1 - \hat{n}_{1\downarrow}\hat{n}_{1\uparrow} - \ldots)(1 - \hat{n}_{1\uparrow}\hat{n}_{1\downarrow} - \ldots) \prod_k \sum_{i,j,\sigma} e^{i(r_i-r_j)\cdot k} |i\sigma, j-\sigma\rangle \tag{A6}$$

Simplifying equations (A5) and (A6),

$$\frac{{}_{\text{NDOC}}\langle\Psi_G|H'_J|\Psi_G\rangle_{\text{NDOC}}}{{}_{\text{NDOC}}\langle\Psi_G|\Psi_G\rangle_{\text{NDOC}}} = \prod_k^{k_F} 2(1-\delta)^2$$

(A7)

[In the previous paper [28] the occupancy condition for the sites was not properly taken into account in the calculations and a slightly different expression was obtained.]

## Appendix B:

$$H_t = \sum_{<m,n>,\sigma}(t_{mn} e^{i\phi_\sigma} C^+_{m\sigma} C_{n\sigma} + t_{nm} e^{-i\phi_\sigma} C^+_{n\sigma} C_{m\sigma}) \tag{B1}$$

Therefore,

$$H_t \, |\Psi_G\rangle_{\text{NDOC}} = \sum_{<m,n>,\sigma}(t_{mn} e^{i\phi_\sigma} C^+_{m\sigma} C_{n\sigma} + t_{nm} e^{-i\phi_\sigma} C^+_{n\sigma} C_{m\sigma})(1-\hat{n}_{1\uparrow}\hat{n}_{1\downarrow} - \hat{n}_{2\uparrow}\hat{n}_{2\downarrow} - \ldots) \prod_k^{k_F} \sum_{i,j} e^{i(r_i-r_j)\cdot k} |i\sigma, j-\sigma\rangle \tag{B2}$$

$$= (t)[\prod_k^{k_F} \sum_{i,j,\sigma} [e^{i\phi_\sigma}\{\delta_{i+1,m}\delta_{i,n}|(i+1)\sigma, j-\sigma\rangle + \delta_{j+1,m}\delta_{j,n}|i\sigma,(j+1)-\sigma\rangle\} + e^{-i\phi_\sigma}\{\ldots\ldots\ldots\}] e^{i(r_i-r_j)\cdot k}$$
$$- \sum_l \prod_k^{k_F}[e^{i\phi_\sigma}\{\delta_{l+1,m}\delta_{l,n}|(l+1)\sigma, l-\sigma\rangle + \delta_{l+1,m}\delta_{l,n}|l\sigma,(l+1)-\sigma\rangle\} + e^{-i\phi_\sigma}\{\ldots\ldots\}]] \tag{B3}$$

where the sum over *l* is carried out involving all the occupied sites $N_l$.

Simplifying equation (B3) and using (A6) we get,

$$T(\phi \neq 0) = (t)\left[\prod_{k,\sigma}^{k_F} \sum_\sigma 4\cos(ka)(1-\delta)^2 \cos(\phi_\sigma) - N_l \prod_{k,\sigma}^{k_F} \sum_\sigma 4\cos(ka)\cos(\phi_\sigma)/N^2\right] \quad (B4)$$